\documentclass[lettersize,journal]{IEEEtran}

\usepackage{etoolbox}
\usepackage{xspace}
\usepackage{amsmath}
\usepackage{amssymb}

\newcommand{\real}{\ensuremath{\mathbb{R}}} %
\newcommand{\abs}[1]{\ensuremath{\left|#1\right|}} %

\newcommand{\setdelimiter}{:}

\newcommand{\setprop}[2]{\left\{#1\setdelimiter#2\right\}} %

\newcommand{\diff}{\ensuremath{\mathrm{d}}}

\newbool{nabla} %
\boolfalse{nabla}

\makeatletter
\DeclareMathOperator{\grad@word}{grad}
\DeclareMathOperator{\rot@word}{rot}
\DeclareMathOperator{\curl@word}{curl}
\DeclareMathOperator{\div@word}{div}

\newcommand{\grad@symbol}{\nabla}
\newcommand{\rot@symbol}{\nabla\times}
\newcommand{\curl@symbol}{\nabla\times}
\newcommand{\div@symbol}{\nabla\cdot}

\newcommand{\Grad}[1][]{\ifstrempty{#1}{\ifbool{nabla}{\grad@symbol}{\grad@word}}{\ifstrequal{#1}{nabla}{\grad@symbol}{\ifstrequal{#1}{word}{\grad@word}{\PackageError{\packagename}{Parameter #1 is not defined for the macro \grad.}{The input parameter #1 that was passed to the macro \grad is not defined. Use none or one of the following input paramters: nabla, word.}}}}}
\newcommand{\Div}[1][]{\ifstrempty{#1}{\ifbool{nabla}{\div@symbol}{\div@word}}{\ifstrequal{#1}{nabla}{\div@symbol}{\ifstrequal{#1}{word}{\div@word}{\PackageError{\packagename}{Parameter #1 is not defined for the macro \div.}{The input parameter #1 that was passed to the macro \div is not defined. Use none or one of the following input paramters: nabla, word.}}}}}
\newcommand{\Rot}[1][]{\ifstrempty{#1}{\ifbool{nabla}{\rot@symbol}{\rot@word}}{\ifstrequal{#1}{nabla}{\rot@symbol}{\ifstrequal{#1}{word}{\rot@word}{\PackageError{\packagename}{Parameter #1 is not defined for the macro \rot.}{The input parameter #1 that was passed to the macro \rot is not defined. Use none or one of the following input paramters: nabla, word.}}}}}
\newcommand{\Curl}[1][]{\ifstrempty{#1}{\ifbool{nabla}{\curl@symbol}{\curl@word}}{\ifstrequal{#1}{nabla}{\curl@symbol}{\ifstrequal{#1}{word}{\curl@word}{\PackageError{\packagename}{Parameter #1 is not defined for the macro \curl.}{The input parameter #1 that was passed to the macro \curl is not defined. Use none or one of the following input paramters: nabla, word.}}}}}

\makeatother

\newcommand{\ds}{\diff s}

\newcommand{\dV}{\diff V}

\newcommand{\vect}[1]{\ensuremath{\vec{#1}}} %

\newcommand{\er}[1][]{\ensuremath{\vect{e}_{r#1}}}

\newcommand{\Ev}{\ensuremath{\vect{E}}} %
\newcommand{\Dv}{\ensuremath{\vect{D}}} %
\newcommand{\Jv}{\ensuremath{\vect{J}}} %
\newcommand{\Av}{\ensuremath{\vect{A}}} %

\newcommand{\mat}[1]{\ensuremath{\mathbf{#1}}} %
\newcommand{\trans}{^{\mkern-1.5mu\mathsf{T}}} %

\usepackage[english]{babel}
\usepackage[babel, autostyle]{csquotes}
\usepackage{xparse}	
\usepackage{ifthen}
\usepackage{tabularray}
\usepackage{acro}
\usepackage[mode=image]{standalone}

\usepackage{subcaption}
\usepackage{ninecolors}
\usepackage{tudacolors}

\usepackage{amsmath}
\usepackage{amssymb}
                   
\usepackage{mathtools}					

\usepackage{siunitx}
\DeclareSIUnit\epszero{\permittivityvacuum}
\DeclareSIUnit\muzero{\permeabilityvacuum}
\DeclareSIUnit\nuzero{\reluctivityvacuum}

\booltrue{nabla}
\usepackage{nicefrac}

\usepackage{tikz}
\usepackage{pgfplots}
\pgfplotsset{compat=1.18}
\usepackage{pgffor}
\usepackage{pgfmath}

\usetikzlibrary{decorations.pathmorphing, decorations.pathreplacing, decorations.shapes,decorations.markings,math,3d,circuits.ee.IEC,calc,arrows.meta, positioning, pgfplots.colormaps,patterns, patterns.meta}
\usepgfplotslibrary{colorbrewer, patchplots}

\tikzset{>=Stealth}
\tikzset{level/.style={%
		execute at begin scope={\pgfonlayer{#1}},
		execute at end scope={\endpgfonlayer}
}}

\pgfplotscreateplotcyclelist{myColorCycleList}{black,TUDa-2b,TUDa-9b,TUDa-3c,TUDa-7b,TUDa-11b,TUDa-10c,TUDa-8c}
\pgfplotsset{cycle list name={myColorCycleList}}
\pgfplotsset{every axis plot/.style={thick,mark=none}}
\pgfplotsset{field plot element/.style={line width=0pt,faceted color=none}}
\pgfplotsset{field plot node/.style={line width=0pt,faceted color=none,shader=interp}}

\usepackage{hyperref}
\hypersetup{breaklinks=true,colorlinks=true,citecolor=blue,hidelinks}%

\newcommand{\vdf}{\vec{\zeta}}

\newcommand{\markconductor}{c}  %
\newcommand{\markinsulation}{i}  %
\newcommand{\markcapacitive}{cap}  %
\newcommand{\markconductive}{c}  %
\newcommand{\markhomogenized}{h}  %

\newcommand{\unitvector}[1]{\vec{e}_{#1}}
\newcommand{\pmat}[1]{\begin{pmatrix}#1\end{pmatrix}}

\newcommand{\dS}{\diff S}

\newcommand{\adof}{\mat{a}}
\newcommand{\udof}{\mat{u}}

\newcommand{\examplefrequencylow}{\qty{20}{\hertz}}
\newcommand{\examplefrequencymed}{\qty{2}{\kilo\hertz}}
\newcommand{\examplefrequencyhigh}{\qty{200}{\kilo\hertz}}

\newcommand{\lcs}[1]{\hat{#1}}  %

\newcommand{\domain}{\Omega}

\newcommand{\fwdomain}{\domain_{\mathrm{fw}}}
\newcommand{\reffwdomain}{\lcs{\domain}_{\mathrm{fw}}}

\newcommand{\slice}{\Gamma}

\newcommand{\crosssection}{S}
\newcommand{\refcrosssection}{\lcs{\crosssection}}

\newcommand{\frequency}{f}
\newcommand{\angularfrequency}{\omega}
\newcommand{\skindepth}{\delta}

\newcommand{\voltage}{V}
\newcommand{\current}{I}
\newcommand{\volfun}{\Phi}
\newcommand{\refvolfun}{\lcs{\volfun}}
\newcommand{\capcurrent}[1][]{\current_{\mathrm{\markcapacitive}#1}}
\newcommand{\condcurrent}[1][]{\current_{\mathrm{\markconductive}#1}}

\newcommand{\thicknessconductor}[1][]{b_{\mathrm{\markconductor}#1}}
\newcommand{\thicknessinsulation}[1][]{b_{\mathrm{\markinsulation}#1}}
\newcommand{\thicknessfoil}{b}
\newcommand{\refthicknessfoil}{\lcs{\thicknessfoil}}
\newcommand{\windingthickness}{w}
\newcommand{\windingheight}{h}
\newcommand{\fillfactor}{\lambda}
\newcommand{\turns}{N}

\newcommand{\distancefrombox}{d}

\newcommand{\yokethickness}{d_{\mathrm{y}}}
\newcommand{\yokeheight}{h_{\mathrm{y}}}
\newcommand{\yokewidth}{w_{\mathrm{y}}}
\newcommand{\airgapthickness}{d_{\mathrm{a}}}
\newcommand{\cornerradius}{r_{\mathrm{c}}}

\newcommand{\reluctivity}{\nu}
\newcommand{\permittivity}{\varepsilon}
\newcommand{\conductivity}{\sigma}
\newcommand{\permeability}{\mu}

\newcommand{\permittivityvacuum}{\permittivity_{0}}
\newcommand{\permeabilityvacuum}{\permeability_{0}}
\newcommand{\reluctivityvacuum}{\reluctivity_{0}}

\newcommand{\reluctivityconductor}{\reluctivity_{\mathrm{\markconductor}}}
\newcommand{\permittivityconductor}{\permittivity_{\mathrm{\markconductor}}}
\newcommand{\conductivityconductor}{\conductivity_{\mathrm{\markconductor}}}
\newcommand{\permeabilityconductor}{\permeability_{\mathrm{\markconductor}}}
\newcommand{\reluctivityinsulation}{\reluctivity_{\mathrm{\markinsulation}}}
\newcommand{\permittivityinsulation}{\permittivity_{\mathrm{\markinsulation}}}
\newcommand{\conductivityinsulation}{\conductivity_{\mathrm{\markinsulation}}}

\newcommand{\permittivityhomogenized}{\permittivity_{\mathrm{\markhomogenized}}}

\newcommand{\permeabilityyoke}{\permeability_{\mathrm{y}}}

\newcommand{\Jvs}{\Jv_{\mathrm{s}}}
\newcommand{\capDv}{\Dv_{\mathrm{\markcapacitive}}}
\newcommand{\meanpotential}[1][]{\phi_{\mathrm{m}#1}}
\newcommand{\refmeanpotential}[1][]{\lcs{\phi}_{\mathrm{m}#1}}
\newcommand{\fwbasis}{p}
\newcommand{\reffwbasis}{\lcs{\fwbasis}}
\newcommand{\edgebasis}{\vec{w}}
\newcommand{\numedgebasis}{N_{\mathrm{a}}}
\newcommand{\numfwbasis}{N_{\mathrm{u}}}

\newcommand{\locala}{\alpha}
\newcommand{\localb}{\beta}
\newcommand{\localc}{\gamma}
\newcommand{\elocala}{\unitvector{\locala}}
\newcommand{\elocalb}{\unitvector{\localb}}
\newcommand{\elocalc}{\unitvector{\localc}}
\newcommand{\domainlocala}{L_{\locala}}
\newcommand{\domainlocalb}{L_{\localb}}
\newcommand{\domainlocalc}{L_{\localc}}
\newcommand{\fwmap}{f}

\newcommand{\stiffnessmatrixsymbol}{K}

\newcommand{\stiffnessmatrix}[1][]{\mat{\stiffnessmatrixsymbol}_{\reluctivity#1}}

\newcommand{\massmatrixsymbol}{M}

\newcommand{\massmatrix}[1][]{\mat{\massmatrixsymbol}_{\conductivity#1}}

\newcommand{\distributionmatrixsymbol}{X}

\newcommand{\distributionmatrix}[1][]{\mat{\distributionmatrixsymbol}_{\conductivity#1}}

\newcommand{\conductancematrixsymbol}{G}

\newcommand{\conductancematrix}[1][]{\mat{\conductancematrixsymbol}_{\conductivity#1}}

\newcommand{\capacitancematrixderivsymbol}{P}

\newcommand{\capacitancematrixderiv}[1][]{\mat{\capacitancematrixderivsymbol}_{\permittivity#1}}

\newcommand{\capacitancematrixnormsymbol}{Q}

\newcommand{\capacitancematrixnorm}[1][]{\mat{\capacitancematrixnormsymbol}_{\permittivity#1}}

\newcommand{\voltagevectorsymbol}{c}

\newcommand{\voltagevector}{\mat{\voltagevectorsymbol}}

\newcommand{\rhsvectorsymbol}{r}

\newcommand{\rhsvector}{\mat{\rhsvectorsymbol}}

\colorlet{accent}{TUDa-10c}
\colorlet{accent2}{azure4}
\newcommand{\accentname}{purple\xspace}
\newcommand{\accenttwoname}{blue\xspace}

\pgfplotsset{fieldlines axis/.style={axis equal image, unbounded coords=jump,xmin=-0.01,xmax=0.02,ymin=-0.01,ymax=0.03,ticks=none,axis lines=none}}

\pgfplotscreateplotcyclelist{currentscyclelist}{black, accent, accent2}
\pgfplotscreateplotcyclelist{viridis eight}{[samples of colormap=8 of viridis]}
\pgfplotsset{currents axis/.style={width=6.5cm, height=3cm, xticklabel=\empty, ylabel={$\current$ [\unit{\ampere}]},cycle list name={currentscyclelist}}}
\pgfplotsset{
	currents axis left/.style={currents axis, anchor=south east,legend style={at={(1.02,0)},anchor=west},xticklabel=\empty,},
	currents axis right/.style={currents axis,anchor=north east,xtick={-1,0,1},	xticklabels={$-\frac{\windingthickness}{2}$,0,$\frac{\windingthickness}{2}$}}}

\NewTblrEnviron{colortblr}
\SetTblrInner[colortblr]{colspec={ll},hlines,vlines,hline{2} = {1pt}, row{odd} = {bg=accent!10!white},
	row{1} = {bg=accent, fg=white}}

\DeclareAcronym{fe}{short={FE}, long={finite element}, list={Finite element}}
\DeclareAcronym{fem}{short = FEM, long = finite element method, list = Finite element method}
\DeclareAcronym{vdf}{short={VDF}, long={voltage distribution function}}
\DeclareAcronym{cdf}{short={CDF}, long={current distribution function}}
\DeclareAcronym{pde}{short = PDE, long = partial differential equation, list=Partial differential equation}
\DeclareAcronym{ode}{short = ODE, long = ordinary differential equation, list=Ordinary differential equation}
\DeclareAcronym{mqs}{short={MQS}, long={magnetoquasistatic}}

\newlength{\figurewidth}
\usepackage[style=ieee,maxnames=1, minnames=1,nohashothers=true,isbn=false,doi=true,url=false,sorting=none]{biblatex}
\usepackage{biblatex-shortfields}
\pgfplotsset{table/search path={figures/data}}
\makeatletter
\def\input@path{{figures/data}}
\makeatother

\begin{document}
\title{Considering Capacitive Effects in Foil Winding Homogenization}%
\author{Jonas Bundschuh, Yvonne Späck-Leigsnering and Herbert De Gersem
\thanks{Manuscript created January, 2024; 
	The work is supported by the German Science Foundation (DFG project 436819664), the Graduate School Computational Engineering at the Technical University of Darmstadt, and the Athene Young Investigator Fellowship of the Technical University of Darmstadt.

	Jonas Bundschuh, Yvonne Späck-Leigsnering and Herbert De Gersem are with the Institute for Accelerator Science and Electromagnetic Fields and the Graduate School of Excellence Computational Engineering at the Technical University of Darmstadt, 64289 Darmstadt, Germany (e-mail: \href{mailto:jonas.bundschuh@tu-darmstadt.de}{jonas.bundschuh@tu-darmstadt.de}; \href{mailto:spaeck@temf.tu-darmstadt.de}{spaeck@temf.tu-darmstadt.de};  \href{mailto:degersem@temf.tu-darmstadt.de}{degersem@temf.tu-darmstadt.de}
}}

\maketitle

\begin{abstract} %
In conventional finite element simulations, foil windings with a thin foil and many turns require many mesh elements. 
This renders models quickly computationally infeasible.
With the use of homogenization approaches, the finite element mesh does not need to resolve the small-scale structure of the foil winding domain. 
Present homogenization approaches take resistive and inductive effects into account. 
With an increase of the operation frequency of foil windings, however, capacitive effects between adjacent turns in the foil winding become relevant. 
This paper presents an extension to the standard foil winding model that covers the capacitive behavior of foil windings. 
\end{abstract}


\section{Introduction}
\IEEEPARstart{T}{he} utilization of foil windings has emerged as a prominent and innovative solution for the design and construction of transformers and inductors \cite{CalderonLopez2019a,Rylko2008a,Ismail2023a,Das_2020aa}. %
Foil windings, characterized by their layered configuration of thin, insulated metal foils, offer distinct advantages over traditional wire-wound counterparts, including enhanced thermal performance \cite{Kazimierczuk_2010aa, Leuenberger_2015aa}, higher fill-factors \cite{Barrios_2015aa,Driesen_2000aa}, and easier construction \cite{De-Gersem_2001aa,Rios_2020aa}.

The foil thickness is very small compared to its other dimensions and to the dimension of the entire foil winding, yielding a multiscale problem.
This comes with the typical drawbacks of the numerical simulation of such problems: If the mesh resolves all small-scale details with a sufficient mesh quality, it becomes quickly very large and the simulation time increases drastically. 
A general remedy is to use multiscale methods, such as the multiscale finite element method \cite{Efendiev_2009aa}, the heterogeneous multiscale method \cite{E_2003aa} or the generalized finite element method \cite{Babuska_2004aa}.
Another possibility is to use a homogenization technique. It searches for effective material parameters to be used on the macro scale simulation of foil windings.

A first conductor model for the homogenization of foil windings was described in \cite{De-Gersem_2001aa} for two-dimensional geometries and in \cite{Geuzaine_2001aa} and \cite{Dular_2002aa} for three-dimensional geometries. 
Furthermore, it was formulated for axisymmetric geometries in \cite{Valdivieso_2021aa}. 
All these formulations have in common that they consider resistive and inductive effects. Capacitive effects, however, are excluded by these formulations.
For applications operated at sufficiently low frequencies, this is a reasonable assumption. 
But at elevated frequencies, capacitive effects can no longer be neglected \cite{Dular2006a,Ioan2000a}, as they can e.g. lead to resonances \cite{DeGreve2013b,Biela2005}.
In power electronics, for instance, the switching frequency is increased to the megahertz range \cite{DeGreve2013a}.

There exist approaches to also consider capacitive effects between windings. 
In \cite{Dular2006a}, the simulation is split into a magnetic model, that calculates the inductive effects, and an electric model, that calculates the capacitive effects afterwards.
Other approaches compute the static capacitances beforehand and use them in an electrical network in a second step \cite{Ioan2000a,Deswal2019a,Chagas2019,Das_2020aa}.
These methods are cumbersome because they split the resistive and inductive effects from the capacitive ones and need multiple simulation runs. 

In this paper, a novel homogenization model for foil windings is presented that considers resistive, inductive and capacitive effects together. It uses the same approximation and ansatz of the standard foil winding model from \cite{De-Gersem_2001aa,Dular_2002aa,Paakkunainen_2024aa} and extends it for capacitive effects.
The structure of this paper is as follows: In the following section, we first define foil windings and introduce the notation. Then, we briefly repeat the standard foil winding model in Sec.~\ref{sec:standard_fw_model} before we present the novel capacitive foil winding model in Sec.~\ref{sec:capacitive_fw_model}. The discretization of this model is the content of Sec.~\ref{sec:discretization}. 
At the end, we present two examples and discuss the simulation results in Sec.~\ref{sec:numerical_examples}.

\section{Definitions}\label{sec:definitions}
A foil is a very thin sheet of metal that is insulated at its surface. Winding the foil around a non-conductive support, creates a foil winding.
$\turns$ denotes the number of turns of the foil winding. 
Furthermore, we use $\windingheight$ for the height and $\thicknessfoil$ for the thickness of the foil. The latter consists of the thickness of the conductor $\thicknessconductor$ and the thickness of the insulation $\thicknessinsulation$, i.e., it is $\thicknessfoil=\thicknessconductor+\thicknessinsulation$.  
The thickness of the foil winding is $\windingthickness=\turns\thicknessfoil$. 
Because often, the relation of thicknesses is needed, we define the fill factor as 
\begin{equation}\label{eq:def_fill_factor}
	\fillfactor := \frac{\thicknessconductor}{\thicknessfoil}\,.
\end{equation}

\begin{figure}
	\centering
	\begin{subfigure}{.55\columnwidth}
		\centering
		\includestandalone[mode=image]{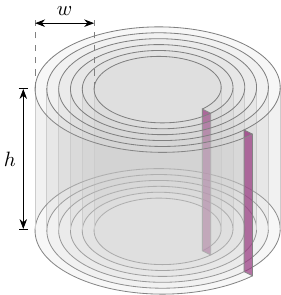}
		\caption{Geometry}
		\label{fig:geometry_foil_winding}
	\end{subfigure}\hfill
	\begin{subfigure}{.35\columnwidth}
		\centering
		\includestandalone[]{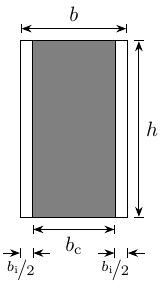}
		\caption{Cross section}
		\label{fig:cross_section_foil}
	\end{subfigure}
	\caption{Schematic representation of a (tube type) foil winding with $\turns=\num{5}$ turns. The geometry with the two terminals highlighted in \accentname is shown in \subref{fig:geometry_foil_winding}. The cross section of the foil with the conductor in gray and the insulation material in white is shown in \subref{fig:cross_section_foil}.}
	\label{fig:tube_type_foil_winding}
\end{figure}

Figure~\ref{fig:tube_type_foil_winding} visualizes a foil winding with $\turns=\num{5}$ turns and the cross section of the foil. 
The foil winding has two terminals, highlighted in \accentname. These are conductively connected through the foil winding, and in the other direction to the powering electrical circuit. 

For the whole paper, the relevant material properties are the reluctivity $\reluctivity$, the conductivity $\conductivity$ and the permittivity $\permittivity$. The reluctivity is the reciprocal value of the permeability $\permeability$. For the two materials that the foil consists of, i.e., the conducting and insulation material, we use the subscripts $\mathrm{\markconductor}$ and $\mathrm{\markinsulation}$, respectively. For example, the reluctivity of the insulation material is denoted by $\reluctivityinsulation$. 

Throughout this paper, it is assumed that the thickness of the foil is much smaller than its height, i.e., $\thicknessfoil\ll\windingheight$. %
Moreover, it is assumed that the foil is thin with respect to the skin depth $\delta=\sqrt{\frac{2}{\angularfrequency\permeabilityconductor\conductivityconductor}}$, with the angular frequency $\angularfrequency=2\pi\frequency$ and the frequency $\frequency$, i.e., it is assumed
\begin{equation}\label{eq:foil_thin_to_skin_depth}
	\thicknessconductor \ll \skindepth\,.
\end{equation}
These two assumptions have the following two consequences for the foil winding models: First, the current density is approximately constant over the thickness of the foil. This will be used later in the derivation of both the standard foil winding model (Sec.~\ref{sec:standard_fw_model}) and the capacitive foil winding model (Sec.~\ref{sec:capacitive_fw_model}). 
Second, the domain of the foil winding is smoothed such that the terminals become a line. 
For the foil winding depicted in Fig.~\ref{fig:tube_type_foil_winding}, the smoothed variant would have a constant inner and outer radius and a constant cross section. 
A further consequence of the smoothing is that there is no conductive connection between the terminals through the foil anymore. 
Instead, the foil winding is modeled as $\turns$ conductive rings. 
The conductive connection of the terminals through the foil winding is modeled by the foil winding model.

\begin{figure}
	\centering
	\includestandalone[mode=image]{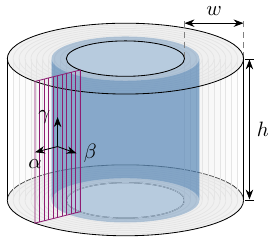}
	\caption{Schematic representation of the foil winding domain $\fwdomain$. The local coordinate system $(\locala,\localb,\localc)$ is shown at one instance of the cross section, highlighted in \accentname. An exemplary representation of $\slice(\locala)$ for one specific $\locala$ is highlighted in \accenttwoname.} 
	\label{fig:foil_winding}
\end{figure}

Similar to \cite{Paakkunainen_2024aa}, we use the local coordinate system $(\locala,\localb,\localc)$ within the \emph{foil winding domain} $\fwdomain\subset\real^3$. The unit vector $\elocala$ is perpendicular to the turns, $\elocalb$ is in the direction of the winding and $\elocalc$ points in the direction of the height of the foil. Consequently, the vectors $\elocala$ and $\elocalc$ span the cross section of the foil winding. Moreover, $\elocalb$ specifies the direction of a resistive current, unaffected from inductive and capacitive effects. Figure~\ref{fig:foil_winding} shows the local coordinate system at one instance of the cross section (highlighted in \accentname) of a foil winding. Note that here and from here on, only the concentric variants are used. 
We define the \emph{foil winding mapping}
\begin{equation}\label{eq:def_fwmap}
	\fwmap:\reffwdomain\mapsto\fwdomain
\end{equation}
as a diffeomorphism, i.e., $\fwmap$ is a bijection and $\fwmap$ and its inverse $\fwmap^{-1}$ are both of class $C^1$. It maps the local coordinates from $\reffwdomain$ to $\fwdomain$. %
The reference domain is defined as
\begin{equation}\label{eq:def_reffwdomain}
	\reffwdomain:=\domainlocala\times\domainlocalb\times\domainlocalc\,,
\end{equation}
with the intervals $\domainlocala,\domainlocalb,\domainlocalc\subset\real$ for the local coordinates. A hat is used whenever a quantity refers to the local coordinate system. 
Based on these definitions, the foil winding domain is parametrized by
\begin{equation}\label{eq:def_fwdomain}
	\fwdomain = \setprop{\fwmap(\locala,\localb,\localc)}{(\locala,\localb,\localc)\in\reffwdomain}\,.
\end{equation}
As indicated in Fig.~\ref{fig:foil_winding}, the foil winding has a constant height $\windingheight$ and constant thickness $\windingthickness$.  %

We define 
\begin{equation}\label{eq:def_slice}
	\slice(\locala) := \setprop{\fwmap(\locala,\localb,\localc)}{\localb\in\domainlocalb,\,\localc\in\domainlocalc}
\end{equation}
to be a slice through the foil winding domain at position $\locala$, as highlighted in Fig.~\ref{fig:foil_winding} in \accenttwoname.

The standard and capacitive foil winding model incorporate two components. The first component consists of finding equivalent material properties, as is also done in classical homogenization approaches \cite{Mackay_2015aa} and computational homogenization \cite{Geers_2010aa}. The second component enforces a particular homogenized condition for the current and is a key point in foil winding homogenization \cite{De-Gersem_2001aa,Dular_2002aa,Paakkunainen_2024aa}.

The homogenized material relations for the reluctivity and conductivity in $\fwdomain$ is found by simple mixing rules \cite{Sihvola_1999aa}.\footnote{For the permittivity, this is not needed, as will be discussed later in Sec.~\ref{sec:capacitive_fw_model}.} The results are tensorial material relations. With respect to the local coordinate system, the respective tensors are diagonal. 
It then is
\begin{align}
	\lcs{\reluctivity} &= \pmat{\reluctivity_\perp&0&0\\0&\reluctivity_\parallel&0\\0&0&\reluctivity_\parallel}\,, & 
	\lcs{\conductivity} &= \pmat{\conductivity_\perp&0&0\\0&\conductivity_\parallel&0\\0&0&\conductivity_\parallel}\,, & 
\intertext{with the reluctivity and conductivity perpendicular to the turns,}
	\reluctivity_\perp &= \fillfactor \reluctivityconductor + \left(1-\fillfactor\right)\reluctivityinsulation\,,&
	\frac{1}{\conductivity_\perp} &= \frac{\fillfactor}{\conductivityconductor} + \frac{1-\fillfactor}{\conductivityinsulation}\,,
\intertext{and the parallel to the turns,}
	\frac{1}{\reluctivity_\parallel} &= \frac{\fillfactor}{\reluctivityconductor} + \frac{1-\fillfactor}{\reluctivityinsulation}\,,&
	\conductivity_\parallel &= \fillfactor\conductivityconductor + \left(1-\fillfactor\right)\conductivityinsulation\,.
\end{align}
These relations follow from the continuity relations of the magnetic field and magnetic flux density and the current density and electrical field strength, respectively. The conductivity of the insulation material is assumed as $\conductivityinsulation=\qty{0}{\siemens\per\meter}$.

\section{Standard foil winding model}\label{sec:standard_fw_model}
The standard foil winding model is derived along the lines of \cite{Paakkunainen_2024aa}, starting from the \ac{mqs} approximation of Maxwell's equations using the $\Av$-$\phi$-formulation, with the vector potential $\Av$ and the scalar potential $\phi$. Thus, the electric field strength is $\Ev = -\partial_t\Av - \Grad\phi$. The \ac{pde} for \ac{mqs} is
\begin{equation}\label{eq:standard_mqs}%
	\Curl\left(\nu\Curl\Av\right) + \sigma\partial_t\Av + \sigma\Grad\phi = \Jvs\,,
\end{equation}
with the source current density $\Jvs$. The computational domain is denoted $\domain\subset\real^3$.

The main idea of the standard foil winding model stems from an approximation of the well-known solid conductor model \cite{Lombard1993a}. With the solid conductor model, it is possible to couple a conducting subdomain in the computational domain to an external circuit. This requires two terminals, between which a voltage $\voltage$ and a current $\current$ can be defined. The external circuit gives a relation between voltage and current and, thus, makes the problem solvable. 

The procedure is as follows: First, an assumption on the scalar potential $\phi$ is made such that the voltage $\voltage$ can be set between the two terminals. Second, an expression for the current $\current$ is derived, explicitly using the assumptions on $\phi$. 
The formulation then comprises a \ac{pde} that contains the voltage and a second equation that determines the current. 

With the winding function of solid conductors from \cite{Schops_2013aa}, which is here denoted as $\vdf$, the ansatz for the scalar potential is $-\Grad\phi = \voltage\vdf$. Since the homogenized foil winding domain consists of $\turns$ conductive turns that are not connected with each other, one could use one solid conductor model for each turn. Assuming that turn $k$ is at the coordinate $\localc_k$, the voltage drop at turn $k$ can be denoted with $\voltage_k$. Since the voltage drop is independent of $\localc$ and $\locala$, as long as still being in turn $k$, it is reasonable to use a function $\refvolfun(\locala)$ to approximate the voltage drop. This function is called \emph{voltage function} and satisfies $\refvolfun(\locala_k) = \voltage_k$. 

The voltage function approximates the voltage $\voltage$ by
\begin{equation}\label{eq:standard_voltage}
	\voltage = \frac{1}{\thicknessfoil} \int_{\Lambda(\localb,\localc)} \volfun\;\ds\,,
\end{equation}
with $\Lambda(\localb,\localc):=\setprop{\fwmap(\locala,\localb,\localc)}{\locala\in\domainlocala}$ and for any $\localb\in\domainlocalb$ and $\localc\in\domainlocalc$. 

To get an expression for the current $\current$, one starts by integrating the current density $\Jv = \sigma\Ev$ over the cross section of a turn of the foil winding. At this point, the assumption $\thicknessfoil\ll\skindepth$ comes into play. With that, the current density can be assumed constant over the thickness of the foil. Thus, the integral is reduced to a one-dimensional integral in $\elocalc$ direction. Also, each turn carries the same current $\current$. Using the winding function $\vdf$, this can be written as
\begin{align}\label{eq:standard_current}	
	\current &= \thicknessfoil \int_{\slice(\locala)} \sigma_\parallel \Ev \cdot\vdf\;\dS\nonumber\\ 
	&= \thicknessfoil \int_{\slice(\locala)} \sigma_\parallel \left(-\partial_t \Av + \volfun\vdf\right) \cdot\vdf\;\dS\,,
\end{align}
which holds for all $\locala\in\domainlocala$.

This completes the standard foil winding model, consisting of \eqref{eq:standard_mqs} with the ansatz for the scalar potential and the expression for the current from \eqref{eq:standard_current}. To summarize, the system of the standard foil winding model reads
\begin{subequations}\label{eq:standard_model}\noeqref{eq:standard_model_mqs}\noeqref{eq:standard_model_curr_cond}%
\begin{align}%
	\Curl\left(\nu\Curl\Av\right) + \sigma\partial_t\Av - \sigma\volfun\vdf &= \Jvs\,, &  &\text{in }\domain\,,\label{eq:standard_model_mqs}\\
	\thicknessfoil\int_{\slice(\locala)} \sigma_\parallel \left(-\partial_t \Av + \volfun\vdf\right) \cdot\vdf\;\dS &= \current\,, & &\locala\in\domainlocala\,,\label{eq:standard_model_curr_cond}
\end{align}
\end{subequations}
with \eqref{eq:standard_voltage} as expression for the voltage and completed with appropriate boundary and initial conditions.

\section{Capacitive foil winding model}\label{sec:capacitive_fw_model}
This section presents a homogenization approach that includes capacitive effects in the foil winding domain.
The general procedure is similar to the standard approach for inductive effects. We start with an ansatz for the voltage and derive an expression for the total current. 

For this section, the assumption that each turn of the foil winding carries the same current has to be dropped because a capacitive current can now flow across the insulation layers. The foil is still thin with respect to the skin depth, i.e., $\thicknessconductor\ll\skindepth$ still holds. 
We further assume that the distance of the foil winding domain to any other conductor in the computational domain is much larger than the distance between two turns in the foil winding, which stems from the high fill factor and very thin insulation layers. %
Thus, capacitive currents outside of the foil winding domain can be neglected. Moreover, because of its structure, capacitive currents inside the foil winding are assumed to be directed only in $\elocala$ direction.

The derivation begins with the same ansatz for the scalar potential as in the standard model, i.e., we set $-\Grad\phi=\volfun\vdf$ with the voltage function $\volfun$ and the winding function for solid conductors $\vdf$ from \cite{Schops_2013aa}. In the reference domain, the voltage function has the form $\refvolfun=\refvolfun(\locala)$. This ansatz models that there is one constant voltage drop over each turn of the foil winding. Consequently, the same equation \eqref{eq:standard_voltage} is used to express the total voltage $\voltage$. 

\begin{figure}
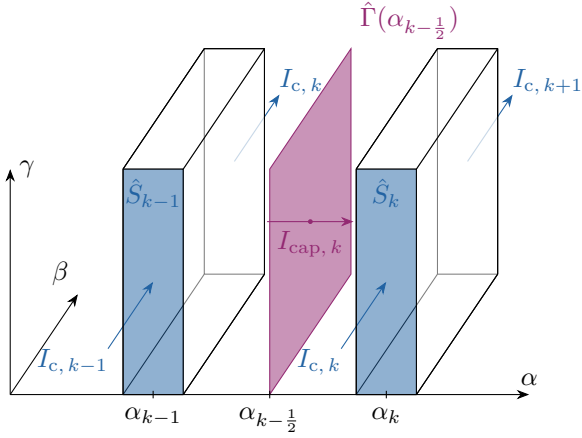

	\centering
	\includestandalone{turns_capacitive}
	\caption{Illustration of two turns, $k-1$ and $k$, of a foil winding in the local coordinate system, located at $\locala_{k-1}$ and $\locala_{k}$. For convenience, the insulation thickness is increased, and the height is reduced. Quantities and surfaces associated with conductive currents are highlighted in \accenttwoname and associated with capacitive currents are highlighted in \accentname. The cross sections of the conducive domains are denoted with $\refcrosssection_{k-1}$ and $\refcrosssection_{k}$, respectively. Conductive currents are denoted with $\condcurrent$ and the capacitive current with $\capcurrent$.}
	\label{fig:turns_capacitive}
\end{figure}

The expression for the total current from \eqref{eq:standard_current} cannot be used because its requirements do not hold anymore. To find a new expression, we consider two turns in a foil winding as depicted in Fig.~\ref{fig:turns_capacitive}. Note that here, for convenience, the insulation thickness has been increased and the height has been reduced. 
Positive currents are in $\elocalb$- and $\elocala$-direction. This means that conductive currents, denoted with $\condcurrent$, are considered positive in $\elocalb$-direction and capacitive currents, denoted with $\capcurrent$, are considered positive in $\elocala$-direction. 

Figure~\ref{fig:turns_capacitive} shows the conductive parts of two turns in the local coordinate system at positions $\locala_{k-1}$ and $\locala_{k}$. 
The current $\condcurrent[,\,k-1]$ enters turn $k-1$ at the front side. 
The current leaving it at the back side is equal to the current entering the adjacent turn in $\elocala$-direction. 
Since, the positions are such that $\locala_k-1<\locala_k$, the current leaving turn $k-1$ at the back side is the current $\condcurrent[,\,k]$, that enters turn $k$ at the front side. 
The conductive currents are determined with
\begin{equation}
	\condcurrent[,\,k] = \int_{\crosssection_{k}} \conductivityconductor\Ev \cdot\elocalb\;\dS\,,
\end{equation}
with the cross section $\crosssection_k$ of the of the conductive part of turn $k$. Between both turns, there flows the capacitive current $\capcurrent[,\,k]$ from turn $k-1$ to turn $k$ over the insulation. This current is given by
\begin{equation}
	\capcurrent[,\,k] = \int_{\slice(\locala_{k-\frac12})} \partial_t\capDv \cdot \elocala\;\dS\,,
\end{equation}
with the (still unknown) displacement field $\capDv$ responsible for the capacitive current.

The global current $\current$ has to flow through the foil winding in $\elocala$- and $\elocalb$-direction. Consequently, it can be approximated with
\begin{equation}\label{eq:total_current_abstract_form}
	\current \approx \condcurrent[,\,k] + \capcurrent[,\,k]
\end{equation}
for each turn $k$. Because two adjacent turns are very close to each other, the capacitive current leaving their separating insulation layer through the surfaces at the top and bottom are neglected. Moreover, because the distance from the turns to any other conductor in the computational domain is much larger than the insulation thickness, the capacitive currents leaving a turn through the top and bottom surface is also neglected. 

The conductive current is approximated with
\begin{equation}\label{eq:conductive_currents}
	\condcurrent[,\,k] \approx \thicknessfoil \int_{\slice(\locala_{k})} \conductivity_\parallel \Ev \cdot\vdf \;\dS\,,
\end{equation}
where $\thicknessconductor\ll\skindepth$ and the properties of the winding function are used. 

\begin{figure}
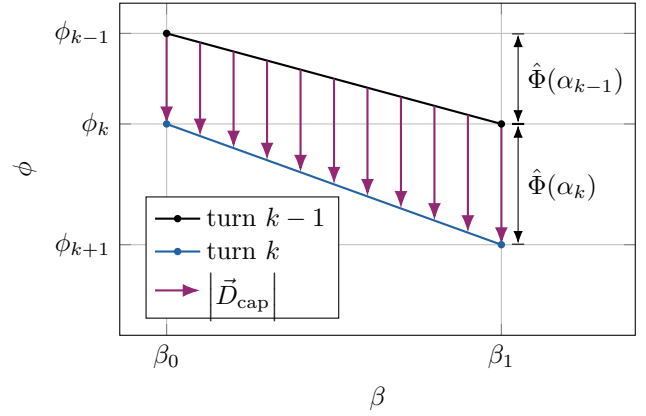

	\centering
	\includestandalone{potential_adjacent_turns}
	\caption{Potential $\phi$ inside turn $k-1$ and $k$ of the foil winding. The potential drop is given by $\volfun(\locala_{k})$. The magnitude of the displacement field $\capDv$ between the two turns is visualized by the length of the \accentname arrows, pointing from the higher to the lower potential of the adjacent turns.}
	\label{fig:pot_adj_turns}
\end{figure}

To approximate the capacitive currents, we first need an expression for the displacement field $\capDv$ in the insulation layers between the turns. For that, we again consider two adjacent turns. Figure \ref{fig:pot_adj_turns} shows the potential $\phi$ in the turns $k-1$ and $k$ along the $\localb$ direction. In turn $k-1$, the potential drops linearly from $\phi_{k-1}$ to $\phi_k$ and in turn $k$ from $\phi_k$ to $\phi_{k+1}$. Note that this potential drop is described by the voltage function $\refvolfun$ at the respective local position. 
The mean potential difference between both turns is given by
\begin{align}%
	\Delta\refmeanpotential[,\,k-1] &= \frac{\phi_{k-1} + \phi_k}{2} - \frac{\phi_k + \phi_{k+1}}{2}\\
	&= \frac12\left(\refvolfun(\locala_{k}) - \refvolfun(\locala_{k-1})\right) + \refvolfun(\locala_{k-1})\\
	&\approx \frac12 \refthicknessfoil \refvolfun'(\locala_{k-1}) + \refvolfun(\locala_{k-1})\,.
\end{align}
Consequently, the displacement field $\capDv$ is approximately
\begin{equation}
	\capDv \approx \permittivityinsulation\frac{1}{\thicknessinsulation} \Delta\meanpotential \elocala = \permittivityinsulation\frac{1}{\thicknessinsulation} \left(\frac12 \thicknessfoil \Grad\volfun\cdot\elocala + \volfun\right)\elocala\,.
\end{equation}
Rewriting the term $\nicefrac{\permittivityinsulation}{\thicknessinsulation}$ with the homogenized permittivity $\permittivityhomogenized = \frac{\permittivityinsulation}{1-\fillfactor}$ as $\nicefrac{\permittivityhomogenized}{\thicknessfoil}$ finally yields the approximate expression for the capacitive current as
\begin{equation}\label{eq:capacitive_currents}
	\capcurrent[,\,k] = \partial_t\int_{\slice(\locala_{k})} \permittivityhomogenized\frac{1}{\thicknessfoil} \left(\frac12 \thicknessfoil \Grad\volfun\cdot\elocala + \volfun\right) \dS\,.
\end{equation}
With the relation from \eqref{eq:total_current_abstract_form}, the expression for the global current is 
\begin{equation}\label{eq:cap_current_expression}
	\current \approx \thicknessfoil \int_{\slice(\locala)} \conductivity_\parallel \Ev \cdot\vdf \;\dS + \partial_t\int_{\slice(\locala)} \permittivityhomogenized \left(\frac12  \Grad\volfun\cdot\elocala + \frac{1}{\thicknessfoil}\volfun\right)\dS
\end{equation}
for all $\locala$ in $\domainlocala$. 

To summarize, the capacitive foil winding formulation reads%
\begin{subequations}\label{eq:capacitive_model}\noeqref{eq:capacitive_model_mqs}\noeqref{eq:capacitive_model_curr_cond}%
\begin{align}%
	\Curl\left(\nu\Curl\Av\right) + \sigma\partial_t\Av - \sigma\volfun\vdf &= \Jvs\,, &  &\text{in }\domain\,,\label{eq:capacitive_model_mqs}\\ 
	\thicknessfoil\int_{\slice(\locala)} \sigma_\parallel \left(-\partial_t \Av + \volfun\vdf\right) \cdot\vdf\;\dS\;+& \nonumber\\
	\partial_t \int_{\slice(\locala)}\permittivityhomogenized \left(\frac12 \Grad\volfun\cdot\elocala + \frac{1}{\thicknessfoil}\volfun\right)\dS &= \current\,, & &\locala\in\domainlocala\,,\label{eq:capacitive_model_curr_cond}
\end{align}
\end{subequations}
with \eqref{eq:standard_voltage} as expression for the voltage and completed with appropriate boundary and initial conditions.

By approximating the displacement field $\capDv$ in the insulation layers between the turns with the potential drop in the turns given by the voltage function $\volfun$, we ended up with expression \eqref{eq:cap_current_expression} for the global current. The derivation does not include the time derivative of the vector potential but only the scalar potential $\phi$. This means that the model includes both inductive and capacitive effects through the magnetic and electric energy, but it excludes radiation effects.
We also restricted the capacitive currents to the interior of the foil winding domain. Therefore, outside of the foil winding domain, the \ac{mqs} formulation is still used.

\section{Discretization}\label{sec:discretization}
In this section, we discretize formulation \eqref{eq:capacitive_model} in space. For the time discretization, a standard method, such as an Euler method can be applied, or the system can be solved in the frequency domain. 
Compared to the standard model formulated in \eqref{eq:standard_model}, the capacitive model has two additional parts in the expression for the global current (see \eqref{eq:capacitive_model_curr_cond}). For the discretization, the same procedure as for the standard model is used. Thus, the discretized version coincides also except for two additional parts. We use the notation of \cite{Paakkunainen_2024aa}.

The vector potential $\Av$ is discretized with \ac{fe} edge shape functions $\edgebasis_j\in\mathbf{H}_0(\text{curl},\domain)$ and the voltage function $\refvolfun$ is discretized on the reference domain with functions $\reffwbasis_j\in H^1(\domainlocala)$. 
Consequently, the voltage function is discretized with $\fwbasis_j\in H^1(\domain)$, where
\begin{equation}
	\fwbasis_j = \begin{cases}
		\reffwbasis_j\circ \fwmap^{-1}_\locala & \text{in } \fwdomain\,,\\
		0& \text{else,}
	\end{cases}
\end{equation}
and where $\fwmap^{-1}_\locala$ denotes the $\locala$ component of the inverse of the foil winding mapping $\fwmap$. Consequently, the discretized vector potential and voltage function are
\begin{align}
	\Av &= \sum_j a_j \edgebasis_j & \text{and}&& \volfun&=\sum_j u_j \fwbasis_j\,.
\end{align}
The coefficients are collected in vectors as $\left[\adof\right]_j = a_j$ and $\left[\udof\right]_j = u_j$.

The Ritz-Galerkin method is applied for discretizing \eqref{eq:capacitive_model}. Multiplication of \eqref{eq:capacitive_model_mqs} with test functions $\edgebasis_i\in\mathbf{H}_0(\text{curl},\domain)$ and integration over $\domain$ yields the matrices $\stiffnessmatrix,\massmatrix\in\real^{\numedgebasis\times\numedgebasis}$ and $\distributionmatrix\in\real^{\numedgebasis\times\numfwbasis}$ and the vector $\rhsvector\in\real^{\numedgebasis}$. Their entries are given by
\begin{align}
	\left[\stiffnessmatrix\right]_{i,j} &= \int_\domain \left(\reluctivity \Curl\edgebasis_j\right) \cdot\Curl\edgebasis_i\;\dV\,,\\
	\left[\massmatrix\right]_{i,j} &= \int_\domain \left(\conductivity \edgebasis_j\right) \cdot\edgebasis_i\;\dV\,,\\
	\left[\distributionmatrix\right]_{i,j} &= \int_\domain \left(\conductivity\fwbasis_j\vdf\right)\cdot\edgebasis_i\;\dV\,,\\
	\left[\rhsvector\right]_{i} &= \int_\domain \Jvs \cdot \edgebasis_i\;\dV\,.
\end{align}
Multiplication of \eqref{eq:capacitive_model_curr_cond} with test functions $\fwbasis_i\in H^1(\domain)$ and integration over $\Lambda(\localb,\localc)$ yields the matrices $\distributionmatrix\trans\in\real^{\numfwbasis\times\numedgebasis}$ and $\conductancematrix,\capacitancematrixderiv,\capacitancematrixnorm\in\real^{\numfwbasis\times\numfwbasis}$ and the vector $\voltagevector\in\real^{\numfwbasis}$. Their entries are given by
\begin{align}
	\left[\conductancematrix\right]_{i,j} &= \int_{\fwdomain} \conductivity \left(\fwbasis_j\vdf\right)\cdot\left(\fwbasis_i\vdf\right)\dV\,,\\
	\left[\capacitancematrixderiv\right]_{i,j} &= \frac{1}{2\thicknessfoil} \int_{\fwdomain} \permittivity \Grad\fwbasis_j\cdot\elocala \fwbasis_i \;\dV\,,\\
	\left[\capacitancematrixnorm\right]_{i,j} &= \frac{1}{\thicknessfoil^2} \int_{\fwdomain} \permittivity \fwbasis_j \fwbasis_i\;\dV\,,\\
	\left[\voltagevector\right]_{i} &= \frac{1}{\thicknessfoil} \int_{\Lambda(\localb,\localc)} \fwbasis_i\;\ds\,.
\end{align}
As for the standard model, the discretization of the expression for the global voltage from \eqref{eq:standard_voltage} is 
\begin{equation}
	\voltage = \voltagevector\trans\udof\,.
\end{equation}

The discretized version of the capacitive foil winding model in space then reads
\begin{subequations}\label{eq:discretized_capacitive_model}
	\begin{align}
		\stiffnessmatrix\adof + \massmatrix\frac{\diff}{\diff t}\adof - \distributionmatrix\udof &= \rhsvector\,,\\
		-\distributionmatrix\trans\frac{\diff}{\diff t}\adof + \conductancematrix\udof + \capacitancematrixderiv\frac{\diff}{\diff t}\udof + \capacitancematrixnorm\frac{\diff}{\diff t}\udof &= \voltagevector\current\,.
	\end{align}
\end{subequations}

\section{Numerical examples}\label{sec:numerical_examples}
\begin{figure}
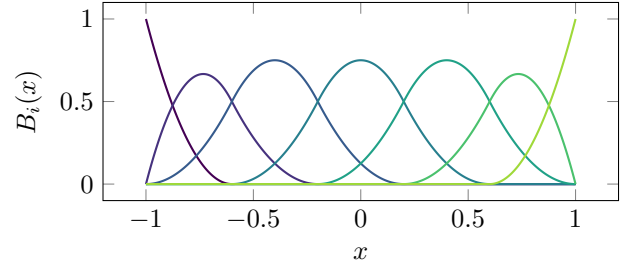

	\centering
	\includestandalone{bsplines}
	\caption{Seven quadratic B-splines, $B_i$, on the interval $[-1,1]$.}
	\label{fig:quadratic_b_splines}
\end{figure}

Two examples are presented in this section. The first one is an academic example to showcase the homogenization technique. It is posed in Cartesian coordinates and shows field lines, the voltage function and the currents in the foil winding. The second one is a realistic example of a pot inductor in cylindrical coordinates. For both examples, \eqref{eq:discretized_capacitive_model} is solved in the frequency domain, i.e., the unknowns are complex numbers and the time derivative becomes a multiplication with $\jmath\omega$. Moreover, quadratic B-splines \cite{de-Boor_2001aa} are used for the functions $\reffwbasis_j$, i.e., for the discretization of the voltage function. 
They are piecewise polynomials of degree two and are continuously differentiable. Because of their compact support and low polynomial degree, they are well suited as a basis for the voltage function. The here used quadratic B-splines are defined with an uniform and open knot vector (see \cite{Hughes_2005aa} and Fig.~\ref{fig:quadratic_b_splines}). 

\subsection{Cartesian example}\label{ssec:example_cartesian}

\begin{figure}
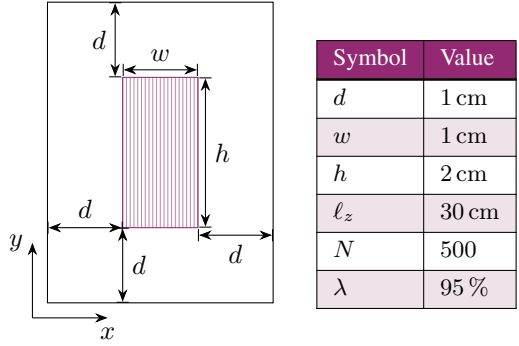

	\centering
	\begin{subfigure}[c]{.5\columnwidth}
		\includestandalone{geometry_example_1}
	\end{subfigure}
	\begin{subfigure}[c]{.45\columnwidth}\small
		\begin{colortblr}{}
			Symbol & Value\\
			$\distancefrombox$ & \qty{1}{\centi\meter}\\
			$\windingthickness$ & \qty{1}{\centi\meter}\\
			$\windingheight$ & \qty{2}{\centi\meter}\\
			$\ell_z$ & \qty{30}{\centi\meter}\\
			$\turns$ & \num{500}\\
			$\fillfactor$ & \qty{95}{\percent}
		\end{colortblr}
	\end{subfigure}
	\caption{Geometry of the first example. The cross section of the foil winding (\accentname) is centered within a box. All flux is confined within the box. The cross sections of the single turns are indicated by the vertical lines (\accentname).}
	\label{fig:geometry_example_cart}
\end{figure}

For the first example, we consider a Cartesian 2D model with length $\ell_z$ of a foil winding. The geometry with its dimensions is shown in Fig.~\ref{fig:geometry_example_cart}. The foil winding is placed in the center of a computational domain filled with air. It has a distance of $\distancefrombox$ from the boundary, where flux wall boundary conditions are applied, i.e., the normal magnetic flux density vanishes at the boundary. The $(x,y,z)$-directions correspond to the $(\locala,\localb,\localc)$-directions, respectively.

In the foil winding, the conducting material has a conductivity of $\conductivityconductor=\qty{60}{\mega\siemens\per\meter}$ and a permittivity of $\permittivityconductor=\permittivityvacuum$. The insulation material has a permittivity of $\permittivityinsulation=\qty{10}{\epszero}$. 

The simulation uses \num{10} quadratic B-splines for the discretization of the voltage function. Furthermore, the mesh counts \num{278912} elements and \num{139737} nodes. The results are presented for the frequencies \examplefrequencylow, \examplefrequencymed\ and \examplefrequencyhigh. 
At these frequencies, the dominating effects are resistive, inductive and capacitive, respectively. 
Each simulation considers a cosine with the respective frequency as current source.

\begin{figure}%
	\centering
	\begin{subfigure}{\columnwidth}
		\centering
		\includestandalone[mode=image]{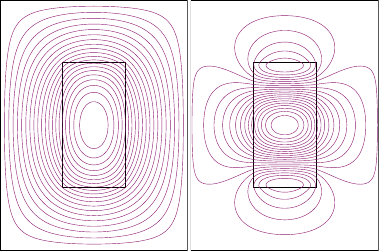}
		\caption{$f=\examplefrequencylow$}
		\label{fig:magnetic_field_lines_low}
	\end{subfigure}
	\begin{subfigure}{\columnwidth}
		\centering
		\includestandalone[mode=image]{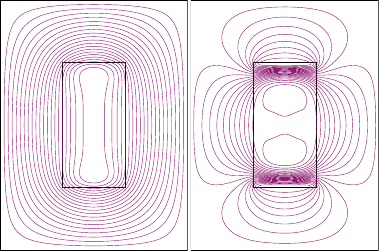}
		\caption{$f=\examplefrequencymed$}
		\label{fig:magnetic_field_lines_med}
	\end{subfigure}
	\begin{subfigure}{\columnwidth}
		\centering
		\includestandalone[mode=image]{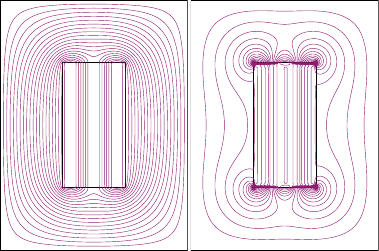}
		\caption{$f=\examplefrequencyhigh$}
		\label{fig:magnetic_field_lines_high}
	\end{subfigure}
	\caption{Real part (left) and imaginary part (right) of the magnetic field lines.}
	\label{fig:magnetic_field_lines}
\end{figure}

The magnetic flux lines are shown in Fig.~\ref{fig:magnetic_field_lines}, where the left side shows the real part and the right side shows the imaginary part. For \examplefrequencylow, the flux lines look similar to those of a solid conductor. 
At \examplefrequencymed, the flux lines get expelled from the foil winding domain in vertical direction, i.e., in the direction of the height of the turns. Also, the (conductive) current mainly flows at the tips of the foil winding, which is at the top and bottom side for this example. 
These two effects become even stronger for \examplefrequencyhigh. Additionally, one observes that the current changes direction within the foil winding. This becomes also visible in the upcoming graphs.

\begin{figure}
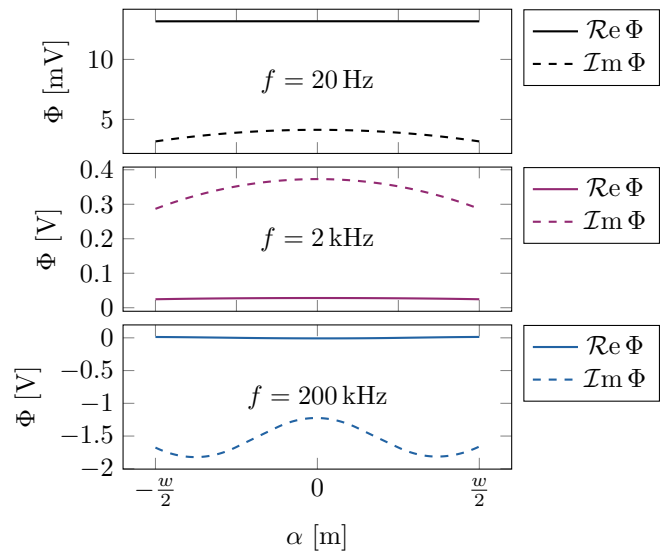

	\centering
	\includestandalone{voltage_function}
	\caption{Voltage function along the $\locala$ direction for different frequencies.}
	\label{fig:voltage_function}
\end{figure}

Next, we take a look at the voltage function $\Phi$. Recall that the voltage function describes the voltage drop over the individual turns of the foil winding. In Fig.~\ref{fig:voltage_function}, $\volfun$ is plotted for the three frequencies over the local coordinate $\locala$, reaching from $-\frac{\windingthickness}{2}$ to $\frac{\windingthickness}{2}$. 
At \examplefrequencylow, the real part of $\volfun$ is greater than the imaginary part. Consequently, the resistive effects dominate. We can already recognize a small inductive effect in the imaginary part. 
In the middle plot, for \examplefrequencymed, the inductive effects are dominant, since the real part is now much smaller than the (positive) imaginary part. By examining the ordinate scale, we observe that the impedance is significantly larger than for \examplefrequencylow, which is the result of the skin depth.
At \examplefrequencyhigh, the imaginary part has switched its sign and is now negative. This indicates that the capacitive effects became dominant (because the source current is a cosine). The shape of the imaginary part has also changed compared to the previous two frequencies. This is covered in the following discussion of the currents.

The conductive current $\condcurrent(\alpha)$ (from \eqref{eq:conductive_currents}) is shown in Fig.~\ref{fig:conductive_currents} and the capacitive current (from \eqref{eq:capacitive_currents}) in Fig.~\ref{fig:capacitive_currents}. As for the voltage function, they are both plotted over the local coordinate $\locala$, reaching from $-\frac{\windingthickness}{2}$ to $\frac{\windingthickness}{2}$. According to \eqref{eq:total_current_abstract_form}, their sum has to be approximately the source current. So the sum of their real parts is approximately equal to \qty{1}{\ampere} and of their imaginary parts to $\qty{0}{\ampere}$, as it can be checked in both figures.
At \examplefrequencylow\ and \examplefrequencymed, the conductive current is dominating. The source current flows mainly conductively through the turns, with a different composition of resistive and eddy currents, depending on the frequency. At \examplefrequencyhigh, however, the capacitive current dominates. It is not at an almost constant value but interacts with the conductive current, resulting in oscillatory shapes. The conductive current changes its sign over the thickness of the foil winding, as already could be observed from Fig.~\ref{fig:magnetic_field_lines_high}. 
\begin{figure}
	\centering
	\includestandalone{conductive_currents}
	\caption{Real and imaginary part of the conductive current $\condcurrent$ in the foil winding domain along the $\locala$ direction for different frequencies.}
	\label{fig:conductive_currents}
\end{figure}
\begin{figure}
	\centering
	\includestandalone{capacitive_currents}
	\caption{Real and imaginary part of the capacitive current $\capcurrent$ in the foil winding domain along the $\locala$ direction for different frequencies.}
	\label{fig:capacitive_currents}
\end{figure}

\subsection{Pot inductor example}\label{ssec:example_pot_inductor}
\begin{figure}
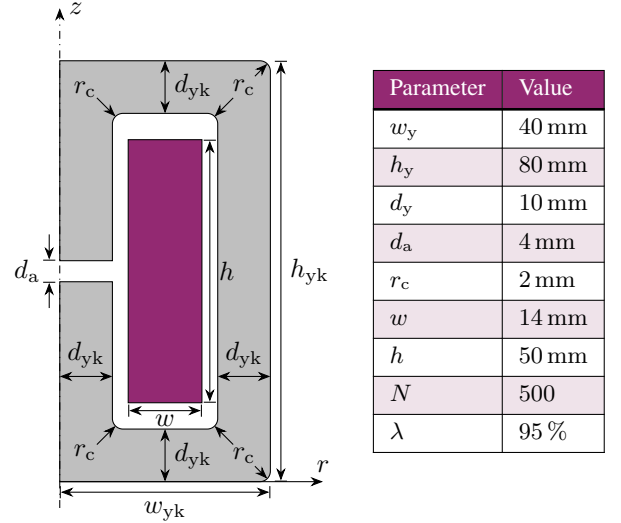

	\centering
	\begin{subfigure}[c]{.6\columnwidth}
		\centering
		\includestandalone{geometry_example_2}
	\end{subfigure}
	\begin{subfigure}[c]{.35\columnwidth}\small
		\begin{colortblr}{baseline=m}
			Parameter & Value\\
			$\yokewidth$ & \qty{40}{\milli\meter}\\
			$\yokeheight$ & \qty{80}{\milli\meter}\\
			$\yokethickness$ & \qty{10}{\milli\meter}\\
			$\airgapthickness$ & \qty{4}{\milli\meter}\\
			$\cornerradius$ & \qty{2}{\milli\meter}\\
			$\windingthickness$ & \qty{14}{\milli\meter}\\
			$\windingheight$ & \qty{50}{\milli\meter}\\
			$\turns$ & \num{500}\\
			$\fillfactor$ & \qty{95}{\percent}
		\end{colortblr}
	\end{subfigure}
	\caption{Geometry of the pot inductor example.}
	\label{fig:geometry_pot_inductor}
\end{figure}

The second example deals with a pot inductor. %
Its geometry with its dimensions is shown in Fig.~\ref{fig:geometry_pot_inductor} in cylindrical coordinates for positive radii. 
The winding region, highlighted in \accentname, is located in the center of the winding window of a cylindrical shaped yoke, highlighted in gray. The yoke has an air gap in the center limb. 

In the winding region, the conducting material has a conductivity of $\conductivityconductor=\qty{60}{\mega\siemens\per\meter}$ and a permittivity of $\permittivityconductor=\permittivityvacuum$. The insulation material has a permittivity of $\permittivityinsulation=\qty{10}{\epszero}$. Moreover, the yoke has a permeability of $\permeabilityyoke = \qty{1000}{\muzero}$ and a vanishing conductivity. 

This example compares the impedance of the pot inductor for the standard and the capacitive foil winding model. A tube type foil winding is considered, i.e., where $\elocala=\er$. The voltage function is discretized with \num{7} quadratic B-splines. 
The simulations are conducted in the frequency domain for frequencies in the range from $\qty{10}{\milli\hertz}$ to $\qty{1}{\mega\hertz}$ and for a mesh with \num{936429} elements and \num{469549} nodes. %

\begin{figure}
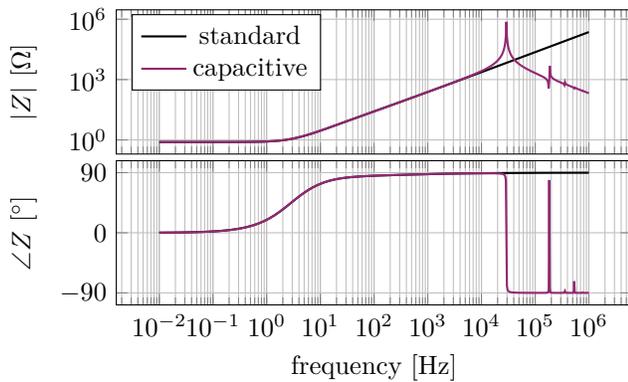

	\centering
	\includestandalone{impedance_sweep_axi}
	\caption{Magnitude and angle of the impedance of the pot inductor over the frequency for the standard (black) and the capacitive (\accentname) foil winding model.}
	\label{fig:impedance_pot_inductor}
\end{figure}

Figure~\ref{fig:impedance_pot_inductor} shows the magnitude and angle of the impedance of the pot inductor. For very low frequencies, the behavior is purely resistive, with a constant magnitude and a phase of \qty{0}{\degree}. Then, both models coincide. With increasing frequency, the magnitude rises as $\abs{Z}\sim\frequency$ and the phase approaches \qty{90}{\degree}. Both models still coincide. At higher frequencies, however, the models give different results. For the chosen problem, this begins around $\frequency=\qty{3}{\kilo\hertz}$. The capacitive model shows the presence of a first resonance frequency at a frequency of approximately $\qty{29}{\kilo\hertz}$ and further resonances at higher frequencies. The phase drops to $\qty{-90}{\degree}$ and the magnitude starts decreasing as $\abs{Z}\sim\nicefrac{1}{\frequency}$, except near resonance frequencies. 
In contrast to that, the standard model keeps showing an inductive behavior of the pot inductor since it does not include the capacitive effects. It is not able to produce resonances or the capacitive behavior above the first resonance frequency.

The computational costs for the capacitive model are only marginally higher than for the standard model.

\section{Conclusion}
In the standard foil winding model, capacitive effects are not considered, which makes it inapplicable for elevated frequencies.  
The presented capacitive foil winding model includes capacitive effects by extending the standard model by two additional terms. 
The displacement current density is expressed in terms of the voltage function. With that, it is not necessary to introduce new degrees of freedom and the overall system size remains the same. 
The first example compared the currents and voltages in the foil winding domain for three exemplary frequencies. At these low, medium and high frequencies, the dominance of resistive, inductive and capacitive effects, respectively, is apparent. 
A comparison of the standard and capacitive foil winding model for a realistic pot inductor showed that the differences between both models are solely in the high frequency regime. Here, only the novel capacitive model shows the presence of the true resonance frequencies and the according capacitive behavior. 
These improved results at high frequencies come with a small additional computational cost of assembling two further matrices. 
\printbibliography

\end{document}